\documentclass[a4paper,10pt]{article}
\usepackage[utf8]{inputenc}
\usepackage{verbatim}
\usepackage[colorlinks]{hyperref}

\title{About a ``nonlocal'' local model considered by L. Vervoort, and the necessity to distinguish locality from Einstein locality}
\author{I. Schmelzer}

\begin{document}

\maketitle

\begin{abstract}
L. Vervoort claims to have found a model which ``can violate the Bell inequality and reproduce the quantum statistics, even if it is based on local dynamics only''. This claim is false. The proposed model contains global elements. The physics behind the model is local, but would not allow the explanation of violations of Bell inequalities for space-like separated events, if superluminal causal influences are forbidden. To use it for this purpose, one has to introduce a preferred frame where information can be send faster than light. 

As a cause of the misunderstanding we identify the unfortunate convention to use ``local'' as a synonym for Einstein-local, so that theories which are local in every physically relevant sense have to be named ``non-local'', and argue that this convention should be abandoned. 
\end{abstract}

%\section{}

In \cite{Vervoort}, L. Vervoort writes:

\begin{quote} [\ldots] a model for the Bell experiment is proposed [which] can violate the Bell inequality and reproduce the quantum statistics, even if it is based on local dynamics only. [\ldots]

Drawing on the essential lessons to be learned from recent fluid-dynamical experiments [\ldots], we showed that such models [\ldots] can violate the Bell inequality and reproduce the quantum correlation of the Bell experiment. [This is] compatible with locality and free will in our model;
superdeterminism does not need to be invoked. Of course, one may well say that such fluid / background models invoke a (harmless) form of ‘delocalized extendedness’ as fluids and fields normally do. But such models do not exhibit the pathological nonlocality that Einstein and Bell sought to exclude. All interactions in a fluid [\ldots] are local in the physically important sense.
\end{quote}

These claims should be rejected.

One problem of the paper is that Vervoort confuses local and global models. So he claims that ``background model M3 does nowhere rely on delocalized or nonlocal interactions; all interactions can be local, as in a fluid''. While this is a correct description for a local description -- which is possible for the process in question -- it is not applicable to his M3 model. This model M3 contains a ``background variable'' $\xi$, which is itself of global nature: It depends on local variables $\lambda_1, \lambda_2$ localized at different places, has no localization itself, thus, is a global variable, and depends (via $\lambda_1, \lambda_2$) on a and b, which is the key difference to the hidden variable $\lambda$ of Bell's theorem ($P(\xi|\lambda_1, \lambda_2)\neq P(\xi)$) which, even if global, can be interpreted as having been local at some distant preparation time. So all information transferred to $\xi$ via some local interaction becomes immediately global. In a truly local model there cannot be such global objects. 

Physically, the variable $\xi$ is justified as describing background information in a stable regime. But such a model is, by construction, a global one, and therefore irrelevant for the discussion of locality. 

But is there, even if one ignores the M3 model as irrelevant for locality, some justification that the physics behind the fluid-dynamical experiments are inherently local? It seems, the physical process considered by Vervoort is a local one in a physically important sense: The system consist of oil droplets which, under some circumstance, bounce rapidly over a vibrating oil film, which results in some 'walking' over the film. This is, in fact, a completely classical system, all objects can be described in a local way, and everything which happens happens with at most the speed of sound of the film. So, the whole system is as non-pathological as any classical theory can be. So can this system be useful for understanding the violation of Bell inequalities?  

This depends on how one tries to use it. The physical system has the speed of the film waves as the limiting speed of information transfer. As explained in the appendix, ``the droplets in the experiments are guided by symmetric waves \ldots. A better approximation is given by Fresnel-Huygens theory, and consists of the superposition of the circular waves created by the droplet at each impact on its trajectory''. The key point to understand why violations of Bell inequalities may be possible is that ``in the stable regime, there is a strong correlation potentially between all system variables''. That means, what is considered is a stable regime -- a regime which presupposes that there has been enough time to establish it, by various interactions. So, this stable regime can tell us nothing about experiments which happen so fast that no wave from one part can reach the other part. Instead, the system is clearly a ``local realistic'' one, so that in this case Bell's inequality can be proven in the usual way. 

Nonetheless, one can try to use this system to explain such violations, if one does not use the speed of the film waves as an analog of the speed of light, but of some new type of information transfer, which happens with a speed much higher than the speed of light, so that even for spacelike separated parts there was enough time using this much faster mechanism to establish the stable regime. Such an attempt would require the introduction of a preferred frame -- the analog of the classical frame in the model -- thus, would be not very popular today. But, nonetheless, this would define a model which is, in any physical sense, local, but nonetheless able to explain violations of Bell inequalities.

Unfortunately, today the naming conventions prevent us from naming such a model ``local'', because actually ``local'' is used synonymously with Einstein-local, and, according to this convention, any model which uses faster-than-light causal influences has to be named ``non-local'', even if the only thing which distinguished it from an Einstein-local theory is that the limiting speed of information transfer is higher. 

As the origin of this confusing convention Bell's original paper \cite{Bell} is not innocent. Bell writes: ``It is the requirement of locality \ldots which creates the essential difficulty'', quotes Einstein \cite{Einstein} with ``But on one supposition we should, in my opinion, absolutely hold fast: The real factual situation of the system $S_2$ is independent of what is done with the system $S_1$, which is spatially separated from the former'' -- above compatible with classical locality -- and connects this later with Lorentz invariance, even if only indirectly: ``the signal involved must propagate instantaneously, so that such a theory could not be Lorentz invariant''. Nonetheless, these statements are unproblematic: The focus of EPR and Bell was incompleteness of quantum theory, and local but not Einstein-local theories cannot exactly recover the quantum predictions too -- if there is some higher limiting speed, there would remain events ``spacelike separated'' even relative to this higher speed where Bell's inequality could be proven, while quantum theory would predict their violation. 

The actual use of ``local'' does not have such a justification. Sometimes it is at least clarified what means ``local'', for example by Gröblacher at al \cite{Zeilinger}: ``According to Bell’s theorem, any theory that is based on the joint assumption of realism and locality (meaning that local events cannot be affected by actions in space-like separated regions) is at variance with certain quantum predictions''. But usually it is omitted. For example, Wittman et al \cite{Zeilinger2} simply write ``Tests of the predictions of quantum mechanics for entangled systems have provided increasing evidence against local realistic theories''. Which they have not, because even in principle this class of experiments can give only lower limits for the maximal speed of information transfer in such theories, as, for example, \cite{Gisin} have shown that this maximal speed has to be $ > 10^4 c$. And this restriction will remain forever, because no test of Bell inequalities can exclude completely that the speed of information transfer is finite. 

One could defend this convention because in science naming conventions are usually irrelevant -- all what matters is that all notions are well-defined. Last but not least, nobody thinks that the colors of QCD have some relation to real colors. But the paper discussed here shows that confusions about the meaning of ``local'' really happen even in published literature. 

Moreover, a naming convention which forces us to name theories which are local in any physically important sense ``non-local'' is not only absurd, but can be even considered as Orwellian.\footnote{To classify the actual convention as ``Orwellian'' is justified not only because it requires to name a local theory non-local. It also shares another important aspect with newspeak -- it leaves some incorrect thoughts without words to talk about then: Indeed, the word ``local'' is the natural word to describe the class of models considered in this paper, with some much higher speed of information transfer in a hidden preferred frame, and to distinguish it from theories with really pathological locality and causality violations. And this is, indeed, a class of theories which is the closest thing to anathema in modern physics.}   

Let's also note that there already exists an adequate and non-misleading name for what is named ``local'' today -- \emph{Einstein-local} -- and  that the these theories without name are important enough to deserve their natural name.\footnote{This class cannot be rejected as so unimportant that it is not even worth to be named adequately. It actually contains viable theories of gravity \cite{Ae} \cite{glet} as well as proposals for high energy physics beyond the standard model \cite{clm} and is clearly important for discussions of realistic or causal interpretations of quantum theory and the violations of Bell's inequalities.} 

So, despite the weak points of the paper, to use vibrating oil droplets on an oil film as a model which allows to explain violations of Bell inequalities by local, but not Einstein-local theories is an interesting idea and deserves further research -- research, which should not be confused by absurd naming conventions which would require to name such a clearly local model non-local.

\end{document}